\documentclass[reprint,nofootinbib,superscriptaddress,amsmath,amssymb,aps]{revtex4-1}
\usepackage{graphicx}
\usepackage{rotating}
\usepackage{dcolumn}
\usepackage{bm}
\usepackage{color}
\usepackage{mathptmx, textcomp}
\usepackage[latin1]{inputenc}
\usepackage{braket}
\usepackage[colorlinks,linkcolor=magenta,anchorcolor=cyan,citecolor=blue]{hyperref}
\usepackage[multiple]{footmisc}
\newcommand{\fig}[1]{Fig.\ref{#1}}
\def\be{\begin{equation}}
\def\ee{\end{equation}}
\def\ba{\begin{eqnarray}}
\def\ea{\end{eqnarray}}

\def\nn{\nonumber}
\def\lf{\left}
\def\rt{\right}

\newcommand{\eq}[1]{(\ref{#1})}

\def\lf{\left}\def\rt{\right}\def\q{\theta} \def\w{\omega}     \def\p {\pi} \def\a {\alpha}  \def\d {\delta} \def\f {\phi}  \def\h {\eta}   \def\l {\lambda} \def\z {\zeta} \def\x {\xi} \def\c {\chi}   \def\m {\mu} \def\pd {\partial}\def\p {\pi} \def \inf {\infty}  \def \e { \varepsilon}
\def\Q{\Theta} \def\W{\Omega}     \def\S {\Sigma}  \def\F {\Phi} \def\G {\Gamma}     \def\grad{\nabla}\def\.{\cdot}
\def\math {\mathcal}
\begin{document}

\title{{ Testing the Weak Cosmic Censorship Conjecture in Lanczos-Lovelock gravity}}
\author{Jie Jiang}
\email{jiejiang@mail.bnu.edu.cn}
\affiliation{Department of Physics, Beijing Normal University, Beijing, 100875, China}
\author{Ming Zhang}
\email{Corresponding author. mingzhang@jxnu.edu.cn}
\affiliation{Department of Physics, Jiangxi Normal University, Nanchang 330022, China}

\date{\today}

\begin{abstract}
In this paper, we test the weak cosmic censorship conjecture (WCCC) in the nearly extremal static charged black holes of Lanczos-Lovelock-Maxwell gravity based on the new version of gedanken experiments proposed by Sorce and Wald. After introducing the null energy condition of the matter fields, we show that the (nearly) extremal black holes can be destroyed under the first-order approximation for the case with $S'(r_h)\leq 0$, where $S(r_h)$ is the entropy of the background black hole geometry in Lanczos-Lovelock gravity. It implies that the WCCC is violated in these situations. For the case with $S'(r_h)>0$, the nearly extremal black holes cannot be overcharged by the new version of the gedanken experiments for both first- and second-order approximations of perturbation. These results indicate that the WCCC is satisfied in the Lanczos-Lovelock gravity with the condition $S'(r_h)>0$. Our work also implies that the WCCC will play a natural role to constrain the Lanczos-Lovelock gravities. Finally, we also show that the destroy condition $S'(r_h)<0$ implies that the nearly extremal black hole is thermodynamically unstable under the first-order approximation.

\end{abstract}
\maketitle
\section{Introduction}

The classical gravitational theories predict the curvature singularity of spacetime. In most cases, it is completely obscured by the event horizon. However, if the event horizon is destroyed, the naked singularity would make the spacetime unpredictable. In response to this problem, Penrose proposed the weak cosmic censorship conjecture (WCCC) \cite{RPenrose} to suppose that the singularity is always hidden inside the event horizon and therefore cannot be detected by the distant observers. Because this conjecture still lacks universal proof until now, it becomes one of the most outstanding unsolved questions in classical gravities. To test this conjecture, Wald devised a gadanken experiment to destroy an extremal Kerr-Newman (KN) black hole by dropping a charged spinning test particle into the event horizon \cite{Wald94}. Their results showed that the extremal KN black hole cannot be destroyed under the first-order approximation and therefore the WCCC is valid. Nevertheless, {  Hubeny showed that overcharging is possible for a nearly extremal Reissner-Nordstrom (RN) black hole in the test-body limit when the second-order effects are neglected \cite{Hubeny}. After that, numerical work showed that the self-force effect may prevent Hubeny-type violations in the nearly extremal RN black holes \cite{SF1} and Kerr black holes \cite{SF2,SF3,SF4,SF5}}. {These results has attracted a lot of researchers to extend the discussion into some other stationary black holes \cite{1,2,3,4,5,B1,B2, B3, B4, B5, B6,B7,B8,B9,B10,B11,B12,B13,B14,B15,B16,B17,Ghosh:2019dzq}.}

{  Recently, Sorce and Wald suggested a new version of the gedanken experiment in which they straightly consider the full dynamical system of gravity and the perturbation matter fields. Then, the self-force effects, finite-size effects, and any other second-order effects are taken into account automatically in the gedanken experiments.} Based on { the covariant phase space formalism \cite{IW,IW2,LW,Wald93}} and the assumption of the null energy condition of the perturbation matter fields, they derived the first- and second-order perturbation inequalities for { the mass and charge of the black hole}. These two inequalities reflected the null energy condition of the perturbation matter fields under the first- and second-order approximation, individually. As a result, they showed that the nearly extremal KN black holes cannot be destroyed by the gedanken experiments and therefore the WCCC is also valid under the second-order approximation.

Most recently, the new version has been extended into various gravitational theories \cite{An:2017phb,Ge:2017vun,Jiang:2019ige,WJ,Jiang:2019vww,Jiang:2019soz,He:2019mqy,JiangZ,JiangG,Jiang:2019soz,Wang:2020vpn}.
All of them showed that the nearly extremal black hole cannot be destroyed in the gedanken experiments under the second-order approximation. However, most of the results are only restricted to the cases where the Einstein gravity couples to some ordinary matter fields, such as scalar field and nonlinear electromagnetic fields. There doesn't exist any evidence to show that the WCCC is valid for any diffeomorphism-covariance gravity. In fact, we will show that there are some violations of the WCCC in some higher-curvature gravitational theories. Then, if we treat the WCCC as a basic principle of the gravitational theories, it can be used to test which higher-curvature gravitational theories are reasonable at the classical level.

As one of the most popular higher-curvature gravity, the Lanczos-Lovelock gravities are the only natural extension of Einstein's theory of gravity in higher dimensions if we insist that the field equations contain only up to the second-order derivative of the metric \cite{Lanczos,Lovelock}. This theory is also free of unphysical ghosts and admits consistent initial value formulation \cite{Lovelock2,Kovacs:2020ywu}. The stability of Lanczos-Lovelock gravity is studied in a series of papers \cite{ST1,ST2,ST3,ST4,ST5,ST6,ST7}. In these papers, it has been shown that the charged black holes are stable under vector-type perturbations, but there are some unstabilities for the scalar-type and tensor-type perturbations. In the following, we will perform the new version of the gadenken experiment to test whether the nearly extremal static charged black hole can be overcharged in the Lanczos-Lovelock gravities and discuss which conditions should be imposed to ensure the validity of the WCCC.

The remaining of this paper is organized as follows. In Sec. \ref{sec2}, we briefly review the Lanczos-Lovelock theory of gravity and discuss the geometry of the static charged spherical black holes under the perturbation of the spherical charged matter fields. We assume that the spacetime finally settles down to a static state and the matter fields satisfy the null energy condition. In Sec. \ref{sec3}, we review the { covariant phase space formalism \cite{IW,IW2,LW,Wald93}} and derive the first two order variational identities of Lanczos-Lovelock gravity. In Sec. \ref{sec4}, after assuming the null energy condition of the matter fields and using the first-order variational identity, we get the first-order perturbation inequality for the mass and charge of the black hole. Based on this inequality, we show that the WCCC no longer holds for all the cases of the Lanczos-Lovelock gravity under the first-order approximation. To be specific, the (nearly) extremal black holes cannot be destroyed under the first-order approximation for the case with $S'(r_h)> 0$, but destroyed for the case with $S'(r_h)\leq0$, where $S(r_h)$ is the entropy of the background black hole geometry in Lanczos-Lovelock gravity. The condition $S'(r_h)<0$ also implies that the nearly extremal black hole is thermodynamically unstable under the first-order approximation. This is the first time to find a violation of the WCCC in gravitational theory after all of the effects are taking into account. In Sec. \ref{sec5}, we derive the second-order perturbation inequality under the optimal condition of the first-order perturbation inequality. Then, we show that the nearly extremal static charged black holes cannot be overcharged in the above perturbation process under the second-order approximation for the case with $S'(r_h)>0$. Finally, the conclusion and discussion are presented in Sec. \ref{sec6}.

\section{Perturbed charged static black hole geometries}\label{sec2}

In this paper, we consider a general Lanczos-Lovelock gravity coupled to a Maxwell field sourced by some extra matter fields in $n$-dimensional spacetime. The Lagrangian $n$-form of this theory is given by{
\ba\begin{aligned}
\bm{L}=\frac{\bm{\epsilon}}{16\p}\left[\sum_{k=1}^{k_\text{max}}\frac{\a_k}{2^k}\d_{a_1 a_2\cdots a_k b_k}^{c_1d_1\cdots c_k d_k}\math{R}_{c_1d_1\cdots c_k d_k}^{a_1b_1\cdots a_k b_k}-F_{ab}F^{ab}\right]+\bm{L}_\text{mt}
\end{aligned}\nn\\\ea
with
\ba\begin{aligned}
\math{R}^{a_1b_1\cdots a_l b_l}_{c_1d_1\cdots c_l d_l}=R^{a_1b_1}_{c_1d_1}\cdots R^{a_lb_l}_{c_ld_l}\,,
\end{aligned}\ea
and $k_\text{max}=[(n-1)/2]$, in which the bracket $[\,\,\bullet\,\,]$ denotes a ceiling function}. Here $\bm{\epsilon}$ is the volume element of this spacetime, $\bm{F}=d\bm{A}$ is the strength of the electromagnetic field $\bm{A}$, $\bm{L}_\text{mt}=\bm{\epsilon}\math{L}_\text{mt}$ is the Lagrangian $n$-form of the extra matter fields which carry the stress energy tensor and electric current
\ba\begin{aligned}
T_{ab}=\frac{2}{\sqrt{-g}}\frac{\d \sqrt{-g}\math{L}_\text{mt}}{\d g^{ab}}\,,\quad j^a=\frac{\d \math{L}_\text{mat}}{\d A_a}\,,
\end{aligned}\ea
the parameter $\a_k$ is some coupling constant with $\a_1=1$ such that it is corresponding to Einstein gravity when the higher curvature corrections are neglected, and
\ba\begin{aligned}
\d^{a_1 \cdots a_k}_{b_1\cdots b_k}&=k!\d_{b_1}^{[a_1}\d_{b_2}^{a_2}\cdots \d_{b_k}^{a_k]}\,.
\end{aligned}\ea
is the generalized Kronecker tensor. The equation of motion can be written as
\ba\begin{aligned}
G_{ab}=8\p \left(T_{ab}^\text{EM}+T_{ab}\right)\,,\\
\grad_a F^{ba}=4\p j^b\,,
\end{aligned}\ea
with{
\ba\begin{aligned}
G^b_a&=-\sum_{k=1}^{k_\text{max}}\frac{\a_k}{2^{k+1}}\d^{b a_1b_1\cdots a_k b_k}_{a c_1 d_1\cdots c_k d_k}\math{R}^{c_1 d_1\cdots c_k d_k}_{a_1 b_1\cdots a_k d_k}\,,\\
T^\text{EM}_{ab}&=\frac{1}{4\p}\left(F_{ac}F_b{}^c-\frac{1}{4}g_{ab}F_{cd}F^{cd}\right)\,.
\end{aligned}\ea
}When the extra matter fields vanish, this theory admits a static charged black hole solution given by \cite{LS1,LS2,LS3}
\ba\begin{aligned}\label{solution}
ds^2&=-f(r)dv^2+2drdv+r^2d\W_{n-2}^2\,,\\
\bm{A}&=-\frac{4\p Q}{(n-3)\W_{n-2}r^{n-3}}dv
\end{aligned}\ea
with the blackening factor $f(r)$ which is solved from the algebraic equation
\ba\begin{aligned}
H\left(\frac{1-f(r)}{r^2}\right)=\frac{16\p M}{\W_{n-2}r^{n-1}}-\frac{32\p^2 Q^2}{(n-3)\W_{n-2}^2r^{2(n-2)}}
\end{aligned}\ea
where $H(x)$ is a polynomial function
\ba\begin{aligned}
H(x)=\sum_{k=0}^{k_\text{max}}\frac{(n-2)!\a_k x^k}{(n-2k-1)!}\,.
\end{aligned}\ea
Here the parameters $M$ and $Q$ denotes the mass and electric charge of this spacetime, separately. {  $\W_{n-2}=2\p^{(n-1)/2}/\G[(n-1)/2]$ is volume of the unit $(n-2)$-dimensional sphere with the line element
\ba\begin{aligned}
d\W_{n-2}^2=d\q_1+\sin^2\q_1d\q_2^2+\cdots +\sin^2\q_1\cdots\sin^2\q_{n-3}d\q_{n-2}^2\,.
\end{aligned}\nn\\\ea}

Next, we will focus on the black hole solutions which contain at least two Killing horizons. The radius $r_h$ of the event horizon is given by the largest root of the blackening factor $f(r)$.  If there does't exist a root of $f(r)$, it describes a naked singularity. For the black hole case, The corresponding temperature, electric potential, and entropy are expressed as
\ba\begin{aligned}
T&=\frac{f'(r_h)}{4\p}\,,\quad \F =\frac{4\p Q}{(n-3)\W_{n-2}r_h^{n-3}}\,,\\
S&=\frac{\W_{n-2}}{4}\sum_{k=1}^{k_\text{max}}\frac{k(n-2)!\a_k r_h^{n-2k}}{(n-2k)!}\,.
\end{aligned}\ea
{ Here the black hole entropy can be obtained from the Wald entropy \cite{IW} or Jacobson-Myers entropy \cite{JM}. One can verify that the first law of this black hole is satisfied, i.e., $\d M=T\d S+\F_H \d Q$.} Moreover, if it also satisfies $f'(r_h)=0$, this solution describes an extremal black hole. The mass and electric charge of the extremal black hole satisfy the constraints,
\ba\begin{aligned}
M&=\frac{\W_{n-2} r_h^{n-3}}{8(n-3)\p}\left[(n-2) r_h^2H(r_h^{-2})-H'(r_h^{-2})\right]\,,\\
Q^2&=\frac{\W_{n-2}^2r_h^{2(n-3)}}{32\p^2}\left[(n-1)r_h^2 H(r_h^{-2})-2H'(r_h^{-2})\right]\,.
\end{aligned}\ea

In the following, we consider a one-parameter family $\f(\l)$ of the field configurations, in which $\f(0)$ is a static charged black hole solution as shown in Eq. \eq{solution} and { $\f(\l)$ with non-zero $\l$ is a dynamic spherically symmetric solution of Lanczos-Lovelock-Maxwell gravity sourced by some spherical charged matter fields in a finite region of the spacetime. }Here we denote $\f(\l)$ to $g_{ab}(\l)$, $\bm{A}(\l)$ and other charged matter fields. The equation of motion is given by
\ba\begin{aligned}
G_{ab}(\l)&=8\p \lf[T_{ab}^\text{EM}(\l)+T_{ab}(\l)\rt]\,,\\
\grad_a^{(\l)} F^{ba}(\l)&=4\p j^b(\l)\,.
\end{aligned}\ea
The above equations of motion indicate that the configuration $\f(\l)$ is treated as a full dynamics system where the self-force effects, finite-size effects, and any other effects are taken into account automatically.  These effects are also showed to prevent Hubeny-type violations \cite{Hubeny} in the nearly extremal RN black holes \cite{SF1} and Kerr black holes \cite{SF2,SF3,SF4,SF5}. When $\l$ is a small parameter, the dynamical process can be regarded as a perturbation.

Generally, we can describe this dynamical geometry by the following line element,
\ba\begin{aligned}\label{vds}
ds^2(\l)=-f(r,v,\l) dv^2+2\m(r, v, \l)dv dr+r^2d\W^2_{n-2},
\end{aligned}\ea
in which $f(r, v, 0)= f(r)$ and $\m(r, v, 0)=1$ for the background spacetime. For later convenience, we choose a gauge condition such that
\ba\begin{aligned}
\left.\x^a A_a(\l)\right|_{r=r_h}=0\,,
\end{aligned}\ea
in which
\ba\begin{aligned}
\x^a=\left(\frac{\pd}{\pd v}\right)^a
\end{aligned}\ea
is a static Killling vector field of the background geometry, $r_h$ is the horizon radius of the background geometry. In the family described by the line element \eq{vds}, $\x^a$ and $r_h$ are independent of the parameter $\l$.
\begin{figure}
\centering
\includegraphics[width=0.48\textwidth]{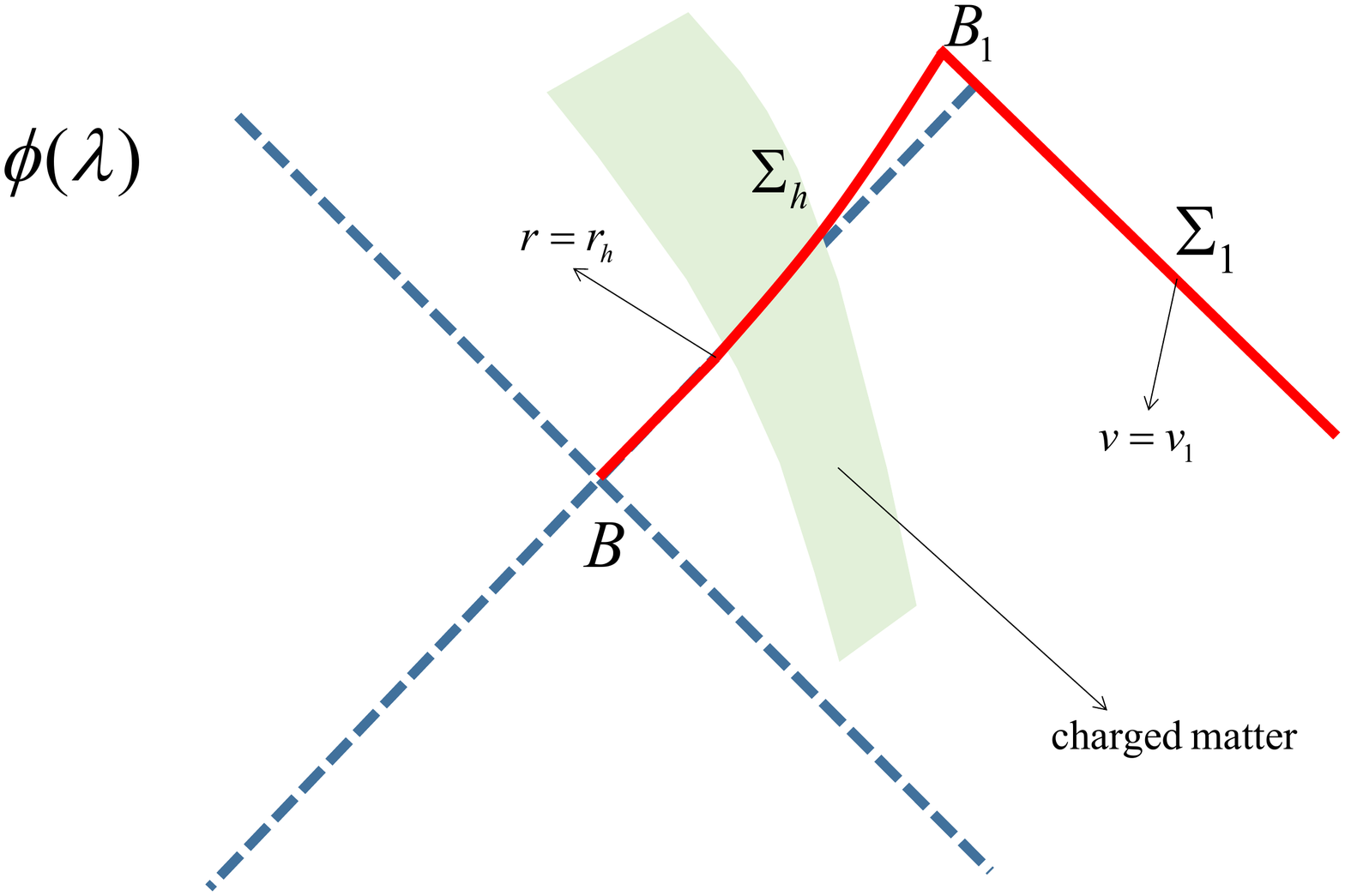}
\caption{A spacetime diagram of the dynamical configuration $\f(\l)$ showing charged matter falling into a nonextremal black hole. $\S_{0}$ is a hypersurface determined by $r=r_h$, where $r_h$ is the horizon radius of the background geometry $\f(0)$. Different from the choice in \cite{SW} where $\math{H}$ is a null hypersurface, $\S_0$ is not a null hypersurface in the configuration $\f(\l)$ with the line element $ds^2(\l)$ because $r_h$ is only the horizon radius of the background geometry.}\label{fig}
\end{figure}

With a similar setup of \cite{SW}, we assume that all the charged matter goes into the black hole through a finite portion of the future horizon (as shown in \fig{fig}) and the spacetime finally settles down to a static state which can also be described by the class of the static charged solution of Lanczos-Lovelock gravity with different electric charge and mass labeled by $\l$, i.e., at asymptotic future, the dynamical fields can be expressed as
\ba\begin{aligned}\label{dsAS}
ds^2(\l)&=-f(r, \l)dv^2+2dr dv+r^2d\W^2_{n-2}\,,\\
\bm{A}&=\frac{4\p Q(\l)}{(n-3)\W_{n-2}}\lf(\frac{1}{r_h^{n-3}}-\frac{1}{r^{n-3}}\rt)dv\,
\end{aligned}\ea
with the blackening factor $f(r,\l)$ which is given by
\ba\begin{aligned}\label{frl}
H\left(\frac{1-f(r, \l)}{r^2}\right)=\frac{16\p M(\l)}{\W_{n-2}r^{n-1}}-\frac{32\p^2 Q^2(\l)}{(n-3)\W_{n-2}r^{2(n-2)}}\,.
\end{aligned}\ea
The above assumption also implies that $T_{ab}(\l)$ and $j^b(\l)$ are vanishing at the sufficiently late times. This is essentially a linear stability assumption as introduced in \cite{SW}. Our spherically perturbation process is belong to the scalar-type perturbations. In Ref. \cite{ST7}, the author studied the stability of the charged Lanczos-Lovelock black holes with the coupling constant $\a_k\geq 0$ for $k\geq 2$. It has been shown that the black holes are unstable for the scalar-type perturbation if the function $2\math{T}'^2-\math{T}\math{T}''$ has negative regions outside the horizon, in which
\ba\begin{aligned}
\math{T}(r)=r^{n-3}H'\left(\frac{1-f(r)}{r^2}\right)\,.
\end{aligned}\ea
However, there still no further investigations to show the stabilities of more general cases. As mentioned in Sec. IV of Ref. \cite{SW}, if the nonextremal black hole were linearly unstable, there would be no need to attempt to overcharge or overspin it in order to destroy it. Therefore, in this paper, we only consider the case where the black hole is linearly stable under the above perturbation process.

\section{covariant phase space formalism and variational identities}\label{sec3}
In this section, we would like to review the covariant phase space formalism \cite{IW,IW2,LW,Wald93} of the Lanczos-Lovelock-Maxwell gravity and derive the first- and second-order variational identities. In the following, we focus on the off-shell variation of Lanczos-Lovelock gravity coupled to the Maxwell field. The Lagrangian $n$-form we considered is expressed as
\ba\begin{aligned}\label{action}
\bm{L}=\frac{\bm{\epsilon}}{16\p}\left[\sum_{k=1}^{k_\text{max}}\frac{\a_k}{2^k}\d_{a_1 a_2\cdots a_k b_k}^{c_1d_1\cdots c_k d_k}\math{R}^{a_1b_1\cdots a_k b_k}_{c_1d_1\cdots c_k d_k}-F_{ab}F^{ab}\right]\,.
\end{aligned}\ea
Taking a variation of the above action, we have
\ba\begin{aligned}\label{var1}
\d \bm{L}=\bm{E}_\f \d\f+d\bm{\Q}(\f,\d\f)\,,
\end{aligned}\ea
with
\ba\begin{aligned}
\bm{E}_\f\d\f&=-\bm{\epsilon}\lf(\frac{1}{2}T^{ab}\d g_{ab}+j^a\d A_a\right)\,,\\
\bm{\Q}(\f,\d\f)&=\bm{\Q}^\text{LL}(\f,\d\f)+\bm{\Q}^\text{EM}(\f,\d\f)\,.
\end{aligned}\ea
in which{
\ba\begin{aligned}\label{Q2}
\bm{\Q}_{a_1\cdots a_{n-1}}^\text{LL}&=-\frac{1}{8\p}\bm{\epsilon}_{ba_1\cdots a_{n-1}}P^{acbd} \grad_a\d g_{cd}\,,\\
\bm{\Q}_{a_1\cdots a_{n-1}}^\text{EM}&=-\frac{1}{4\p}\bm{\epsilon}_{ba_1\cdots a_{n-1}}F^{bc}\d A_c\,,
\end{aligned}\ea
}are the gravitational part and electromagnetic part of the symplectic potential $\bm{\Q}(\f,\d\f)$. Here we have denoted
\ba\begin{aligned}
P_{ab}^{cd}=\frac{1}{16\p}\sum^{k_\text{max}}_{k=1}\frac{k \a_k}{2^k}\d_{ab a_2 b_2\cdots a_k b_k}^{cd c_2 d_2 \cdots c_k d_k}\math{R}^{ab a_2 b_2\cdots a_k b_k}_{cd c_2 d_2 \cdots c_k d_k}\,.
\end{aligned}\ea
Utilizing the symplectic potential, we can define the symplectic current as
\ba\begin{aligned}
\w(\f,\d_1\f,\d_2\f)=\d_1\bm{\Q}(\f,\d_2\f)-\d_2\bm{\Q}(\f,\d_1\f)\,,
\end{aligned}\ea
and it can also be written as
\ba\begin{aligned}
\bm{\w}(\f,\d_1\f,\d_2\f)=\bm{\w}^\text{LL}(\f,\d_1\f,\d_2\f)+\bm{\w}^\text{EM}(\f,\d_1\f,\d_2\f)
\end{aligned}\nn\\\ea
with{
\ba\begin{aligned}\label{w2}
\bm{\w}_{a_1\cdots a_{n-1}}^\text{LL}(\f,\d_1\f,\d_2\f)&=\frac{1}{8\p}\lf[\d_2\lf(\bm{\epsilon}_{ba_1\cdots a_{n-1}}P^{acbd}\rt)\grad_a\d_1 g_{cd}\right.\\
&\left.-\d_1\lf(\bm{\epsilon}_{ba_1\cdots a_{n-1}}P^{acbd}\rt) \grad_a\d_2g_{cd}\right]\\
\bm{\w}_{a_1\cdots a_{n-1}}^\text{EM}(\f,\d_1\f,\d_2\f)&=-\frac{1}{4\p}\left[\d_1(\bm{\epsilon}_{ba_1\cdots a_{n-1}}F^{bc})\d_2 A_c\right.\\
&\left.-\d_2(\bm{\epsilon}_{ba_1\cdots a_{n-1}}F^{bc})\d_1 A_c\right]\,.
\end{aligned}\ea
}Based on the above expressions, we can define a Noether current $(n-1)$-form corresponding to the vector field $\z^a$ as
\ba\begin{aligned}\label{J1}
\bm{J}_\z=\bm{\Q}(\f,\math{L}_\z\f)-\z\.\bm{L}\,,
\end{aligned}\ea
{ where we have denoted
\ba\begin{aligned}
(\z\. \bm{\h})_{a_2\cdots a_{l}}=\z^b \bm{\h}_{b a_2\cdots a_l}
\end{aligned}\ea
for any $l$-form $\bm{\h}$. From the calculation in \cite{IW2},} this current can also be expressed as
\ba\begin{aligned}\label{J2}
\bm{J}_\z=\bm{C}_\z+d\bm{Q}_\z\,,
\end{aligned}\ea
in which
\ba\begin{aligned}
(\bm{C}_\z)_{a_2\cdots a_{n-1}}&=\bm{\epsilon}_{ba_2\cdots a_{n-1}}(T_a{}^b+A_aj^b)\z^a\,,\\
\bm{Q}_\z&=\bm{Q}_\z^\text{LL}+\bm{Q}_\z^\text{EM}\,
\end{aligned}\ea
with
\ba\begin{aligned}\label{Q22}
(\bm{Q}_\z^\text{LL})_{a_1\cdots a_{n-2}}&=-\frac{1}{16\p}\bm{\epsilon}_{aba_1\cdots a_{n-2}}P^{abcd}\grad_c{ \z_d}\,,\\
(\bm{Q}_\z^\text{EM})_{a_1\cdots a_{n-2}}&=-\frac{1}{8\p}\bm{\epsilon}_{aba_1\cdots a_{n-2}}F^{ab}A_c\z^c
\end{aligned}\ea
are the constraint and Noether charge of this gravitational theory, separately. Next, we consider the static Killing vector $\x^a=(\pd/\pd v)^a$ of the background geometry. Using the above expressions as well as the fact that $T_{ab}=j^b=\math{L}_\x\f=0$ for the background fields, the first- and second-order variational identities can be derived and expressed as\cite{SW}
\ba\begin{aligned}\label{varid}
d[\d\bm{Q}_\x-\x\.\bm{\Q}(\f,\d\f)]&+\d \bm{C}_\x=0\,,\\
d[\d^2\bm{Q}_\x-\x\.\d\bm{\Q}(\f,\d\f)]&=\bm{\w}\lf(\f,\d\f,\math{L}_\x\d\f\rt)\\
&-\x\.\d\bm{E}_\f\d\f-\d^2 \bm{C}_\x\,,
\end{aligned}\ea
{ where we have denoted $\math{L}_\z$ to the Lie derivative with respect to the vector field $\z^a$.}

\section{Gedanken experiments under the first-order approximation}\label{sec4}
Now we shall investigate whether the static charged black holes of the Lanczos-Lovelock gravity can be overcharged in the above physical process. Because of the assumption that the spacetime finally settles down to a static state, it is equivalent to checking whether the spacetime geometry at asymptotic future also describes a black hole. Therefore, we define a function
\ba
h(\l)=f(r_m(\l),\l)
\ea
to describe the minimal value of the blackening factor in the asymptotic future. Here $r_m(\l)$ is the minimal radius of the blackening factor $f(r, \l)$, and it can be obtained by
\ba\begin{aligned}\label{fprm}
f'(r_m(\l),\l)=0\,.
\end{aligned}\ea
The WCCC is violated if $h(\l)>0$. From the above expressions, we can see that if the background spacetime is not a nearly extremal black hole (or extremal black hole), the zero-order perturbation will be positive and the higher-order corrections can be neglected. Then, the black hole cannot be overcharged. Therefore, in order to destroy the event horizon, the background spacetime should be assumed as a nearly extremal black hole. For the nearly extremal static charged black holes, we can define a small parameter $\e$ such that $r_m=(1-\e) r_h$ and choose it to be the same order of $\l$. { In the one-parameter family $\f(\l)$, the $p$th-order variation of any quantity $\c(r, v, \q_1, \cdots, \q_{n-2}, \l)$ is defined by
\ba\begin{aligned}
\d^p \c(r, v, \cdots, \l)\equiv \left.\frac{\pd^p \c(r, v, \cdots, \l)}{\pd \l^p}\right|_{\l=0}\,,
\end{aligned}\ea
which implies that we fix the coordinates $\{v, r, \theta_1, \cdots\}$ under the variation. The variation of the mass and charge are given by $\d M=M'(0), \d^2 M=M''(0)$ and $\d Q=Q'(0), \d^2 Q=Q''(0)$.} Then, under the first-order approximation of perturbation, we have
{ \ba\begin{aligned}
h(\l)&=\l \d f(r_m)+O(\l^2)\\
&=\l \d f(r_h)+O(\l^2, \l \e)
\end{aligned}\ea}
{ where we have denoted
\ba
\d f(r_{h/m})=\left.\frac{\pd f(r, \l)}{\pd \l}\right|_{\l=0,r=r_{h/m}}\,,
\ea
and $O(x)$ represents the same order or higher order infinitesimal quantity of $x$.} From Eq. \eq{frl}, we can further obtain
\ba\begin{aligned}
h(\l)&= -\frac{16\p \l}{\W_{n-2}r^{n-3}_hH'(r^{-2}_h)}\left(\d M-\F \d Q\right)+O(\l^2, \l \e)\\
&=-\frac{4\p \l}{S'(r_h)}\left(\d M-\F \d Q\right)+O(\l^2, \l \e)\,.
\end{aligned}\ea

Next, we would like to utilize the first-order variational identity and null energy condition to judge the sign of $h(\l)$ at the first-order approximation of perturbation. To do this, we introduce a hypersurface { $\S=\S_1\cup \S_h$ as shown in the \fig{fig}}, in which { $\S_h$} is a portion of the hypersurface $r=r_h$ connecting the bifurcation surface $B$ and a cross-section $B_1$ at sufficiently late times, and $\S_1$ is a time-slice connecting the cross-section $B_1$ and infinity $S_\inf$ at asymptotic future. According to the setup of this collision process, the dynamical fields on the hypersurface
$\S_1$ can be expressed in Eq. \eq{dsAS}.

Taking an integration of the first-order variational identity in Eq. \eq{varid} on the hypersurface $\S$, we can further obtain
\ba\begin{aligned}\label{var11}
\int_{S_\inf}\lf[\d\bm{Q}_\x-\x\.\bm{\Q}(\f,\d\f)\rt]+\int_{\S_h}\d \bm{C}_\x=0\,.
\end{aligned}\ea
Here we have used the assumption that the perturbation vanishes on the bifurcation surface $B$. { The second term only depends on $\S_h$ because $\d \bm{C}_\x$ vanishes on $\S_1$ by the assumption that there are no sources outside the hypersurface $r=r_h$ at late times. Using the explicit expressions of the dynamical fields in \eq{dsAS} at late times, we can further obtain the gravitational part of the first term at the left-hand side and it is given by
\ba\begin{aligned}
\int_{S_\inf}\left[\d\bm{Q}_\x^\text{LL}-\x\.\bm{\Q}^\text{LL}(\f,\d\f)\right]=\d M\,.
\end{aligned}\ea
For the electromagnetic part, a straightforward calculation gives
\ba\begin{aligned}\label{QEM}
&\int_{S_\inf} \bm{Q}_\x^\text{EM}(\l)=-\frac{4\p Q^2(\l)}{(n-3)\W_{n-2}r_h^{n-3}}\,,\\
&\int_{S_{\inf}} \x\.\bm{\Q}^\text{EM}(\f(\l),\f'(\l))=-\frac{4\p Q(\l)Q'(\l)}{(n-3)\W_{n-2}r_h^{n-3}}\,.
\end{aligned}\ea
Using the above results, we can further obtain
\ba\begin{aligned}
\int_{S_\inf}\left[\d\bm{Q}_\x^\text{EM}-\x\.\bm{\Q}^\text{EM}(\f,\d\f)\right]=-\F_H\d Q\,.
\end{aligned}\ea
Summing these results, the identity \eq{var11} becomes
\ba\begin{aligned}\label{ineq1}
\d M-\F_H\d Q&=-\int_{\S_h}\d \bm{C}_\x= \int_{\S_h}\bm{\tilde{\epsilon}} \d T_v^r\,.
\end{aligned}\ea
where we have used the assumption that $j^b=T_{ab}=0$ in the background configuration as well as the gauge choice of the electromagnetic field such that $A_a(\l)\x^a=0$ on $\S_h$. Here we have denoted $X_v^r=X_a^b\x^a (dr)_b$ and $X_{rr}=X_{ab}(\pd/\pd r)^a (\pd/\pd r)^b$ to the components of the tensor $X_{ab}$, and $\tilde{\bm{\epsilon}}$ is the volume element on $\S_h$, which is defined by $\tilde{\bm{\epsilon}}=h(r) dv\wedge \hat{\bm{\epsilon}}$ with
\ba\begin{aligned}
\hat{\bm{\epsilon}}=r^{n-2}\left[\prod_{i=1}^{n-3}\sin^{n-2-i}\q_i\right]d\q_1\wedge \cdots \wedge d\q_{n-2}\,,
\end{aligned}\ea}

In the following, we shall connect the above result to the null energy condition of the matter fields, which state that $T_{ab}(\l)k^a(\l)k^b(\l)\geq 0 $ for any { future-pointing} null vector field $k^a(\l)$. In this paper, we choose the null vector field as
\begin{equation}
k^a (\lambda) = \x^a + \frac{f(r, v, \l)}{2\m (r, v, \l)} \left(\frac{\pd}{\pd r}\right)^a\,.
\end{equation}
{ Using the expression
\ba\begin{aligned}
(dr)^a=\frac{1}{h(r,v,\l)}\left[\x^a+\frac{f(r,v,\l)}{\m(r,v,\l)}\left(\frac{\pd}{\pd r}\right)^a\right]\,.
\end{aligned}\ea
It is not hard to verify that
\ba\begin{aligned}\label{TTNull}
T_{ab}(\l)k^a(\l)k^b(\l)=h(r,v,\l)T_v^r(\l)+\frac{f^2(r,v,\l)}{4\m^2(r,v,\l)}T_{rr}(\l)\,,
\end{aligned}\nn\\\ea
Considering the fact that $f(r_h, v, 0)=f(r_h)=0$, $T_v^r(0)=0$ and $\m(r, v, 0)=1$, the null energy condition implies
\ba\begin{aligned}
T_{ab}(\l)k^a(\l)k^b(\l)= \l\d T_v^r+O(\l^2)\geq 0
\end{aligned}\ea
on the hypersurface $r=r_h$. Combing the first-order variational identity \eq{ineq1}, we have}
\ba\begin{aligned}
\d M-\F_H\d Q&\geq O(\l)\,.
\end{aligned}\ea

From the above results, we can see that the sign of $h(\l)$ under the first-order approximation of perturbation is dependent on the sign of $S'(r_h)$. If $S'(r_h)<0$, we have $h(\l)\geq 0\,.$ This means that the black hole { is} overcharged after the perturbation matter fields drop into the event horizon and therefore the WCCC will be violated. Moreover, because the black hole can be destroyed in the first order, it is not necessary to consider the higher-order approximation. For the case with $S'(r_h)=0$, from the expression of the line element at asymptotic future in \eq{dsAS}, we can see that this case can only occur when $\d M=\F_H\d Q$. However, we can see that if it is this situation, there doesn't exist any constraint on $\d f(r_h)$, and therefore the WCCC can be violated.

If $S'(r_h)>0$, we have $h(\l)\leq 0$, which implies that the black hole cannot be overcharged under the first-order approximation of perturbation. However, there is an optimal condition such that $h(\l)=0$ under the first-order approximation, i.e., the first-order perturbation inequality is saturated and therefore { we have $\d T_v^r=0$ on $\S_h$.} In this condition, because $h(\l)$ vanishes at first-order, we need to consider the second-order approximation of $h(\l)$ to judge its sign.

\section{Gedanken experiments under the second-order approximation}\label{sec5}
As mentioned in the last section, in the following, we evaluate the value of $h(\l)$ at second order for the case with $S'(r_h)>0$ under the optimal condition of the first-order perturbation inequality. First of all, we would like to derive the second-order perturbation inequality in this situation. { Integration of the second-order variational identity \eq{varid} on $\S$ yields
\ba\begin{aligned}\label{eq2}
&\int_{S_\inf}\left[\d^2 \bm{Q}_\x-\x\.\d\bm{\Q}(\f,\d\f)\right]\\
&=-\int_{\S_h}\x\.\d\bm{E}_\f\d\f-\int_{\S_h}\d^2\bm{C}_\x+\math{E}_{\S}(\f,\d\f)
\end{aligned}\ea
where we denote
\ba\begin{aligned}
\math{E}_{\S}(\f, \d\f)=\int_{\S}\bm{\w}(\f,\d\f,\math{L}_\x\d\f)\,.
\end{aligned}\ea
Here the first two terms at the right-hand side only depend on $\S_h$ as $T_{ab}(\l)=j^a(\l)=0$ on $\S_1$ from \eq{dsAS}. Moreover, because $\x^a=(\pd/\pd v)^a$ is a tangent vector on $\S_h$ (i.e., $r=r_h$), the first term at the right-side vanishes. Since we have chosen a gauge to fix the coordinate $\{v, r, \cdots\}$ in the variation, from the late-time geometry \eq{dsAS}, we can see that $\math{L}_\x g_{ab}(\l)=0$ on $\S_1$, which implies that $\math{L}_{\x}\d g_{ab}=0$ on $\S_1$. That is to say, the last term only depends on the hypersurface $\S_h$.
Using the explicit expressions of the late-time dynamical fields in \eq{dsAS}, the gravitational part at the right-hand side gives
\ba\begin{aligned}
\int_{S_\inf}\left[\d^2 \bm{Q}_\x^\text{grav}-\x\.\d\bm{\Q}^\text{grav}(\f,\d\f)\right]=\d^2 M\,.
\end{aligned}\ea
For the electromagnetic part, from Eq. \eq{QEM}, a straightforward calculation gives
\ba\begin{aligned}
&\int_{S_\inf}\left[\d^2 \bm{Q}_\x^\text{EM}-\x\.\d\bm{\Q}^\text{EM}(\f,\d\f)\right]\\
&=\left[\int_{S_\inf} \frac{\pd^2\bm{Q}_\x^\text{EM}(\l)}{\pd\l^2}-\x\.\frac{\pd \bm{\Q}^\text{EM}(\f(\l),\f'(\l))}{\pd\l}\right]_{\l=0}\\
&=-\F_H\d^2 Q-\frac{4\p \d Q^2}{(n-3)\W_{n-2}r_h^{n-3}}\,.
\end{aligned}\ea
Combining the above results, we have
\ba\begin{aligned}\label{varid21}
\d^2M-\F_H\d^2Q-\frac{4\p\d Q^2}{(n-3)\W_{n-2}r^{n-3}_h}&=\math{E}_h-\int_{\S_h}\d^2\bm{C}_\x\\
&=\int_{\S_h}\bm{\tilde{\epsilon}}\d^2T_v^r+\math{E}_h\,,
\end{aligned}\ea
in which we have denoted
\ba\begin{aligned}
\math{E}_h=\int_{\S_h}\bm{\w}(\f,\d\f,\math{L}_\x\d\f)\,,
\end{aligned}\ea
and used the optimal condition $\d T_v^r=0$ of the first-order perturbation inequality as well as the gauge condition $A_a(\l)\x^a=0$ on $\S_h$. Next, we are going to connect the above results to the null energy condition under the first-order optimal condition $\left.\d T_v^r\right|_{r_h}=0$. From Eq. \eq{TTNull}, we can further obtain
\ba\begin{aligned}\label{NEC2}
T_{ab}(\l)k^a(\l)k^b(\l)=\frac{\l^2}{2}\d^2 T_v^r+O(\l^3)\geq 0\,,
\end{aligned}\ea
on the hypersurface $\S_h$. Combing the second-order variational identity, the null energy condition implies that
\ba\begin{aligned}\label{varid22}
\d^2M-\F_H\d^2Q-\frac{4\p\d Q^2}{(n-3)\W_{n-2}r^{n-3}_h}\geq \math{E}_h+O(\l)\,,
\end{aligned}\ea

Finally, we would like to evaluate the quantity $\math{E}_h$, which can be divided into the gravity part and electromagnetic part, separately. For the electromagnetic part, with a same calculation of Eq. (106) in \cite{IW}, using the gauge condition $\x^a \d A_a=0$ on $\S_h$, it is not difficult to obtain
\ba\begin{aligned}
\int_{\S_0}\bm{\w}^\text{EM}(\f,\d\f,\math{L}_\x\d\f)&=-\frac{1}{2\p}\int_{\S_h}\bm{\epsilon}_{aa_1\cdots a_{n-1}}\x^b\d F^{ac}\d F_{bc}\\
=&\frac{1}{2\p}\int_{\S_h}\bm{\tilde{\epsilon}}\d F^{ac}\d F_{bc}(dr)_a\x^b\,.
\end{aligned}\ea
From the expression of the stress-energy tensor for the electromagnetic field, we have
\ba\begin{aligned}
&\d^2 (T^\text{EM})^a_b (dr)_a \x^b=\frac{1}{2\p}\d F^{ac}\d F_{bc}(dr)_a\x^b\\
&+\frac{1}{4\p}\d^2 F^{ac}(dr)_aF_{bc}\x^b+\frac{1}{4\p} F^{ac}(dr)_a\x^b\d^2 F_{bc}\,.
\end{aligned}\ea
Using the explicit expression of the background configuration, we can see that $\x^bF_{bc}\propto (dr)_c$ and $F^{ac}(dr)_a\propto \x^c$, which implies that the last two terms vanish. Then, we have
\ba\begin{aligned}
\int_{\S_0}\bm{\w}^\text{EM}(\f,\d\f,\math{L}_\x\d\f)
=\int_{\S_h}\bm{\tilde{\epsilon}}\d^2(T^\text{EM})_v^r\,.
\end{aligned}\ea
Using the explicit expression of the background configuration, we can further find that $(T^\text{EM})_v^r=0$ in the background geometry.
Considering the gauge condition $\x^aA_a(\l)=0$ on $\S_h$, we can also obtain $\d(T^\text{EM})_v^r=0$ on $\S_h$. Combining these results, with a similar calculation to Eq. \eq{NEC2}, the null energy condition of the electromagnetic field implies
\ba\begin{aligned}
\int_{\S_0}\bm{\w}^\text{EM}(\f,\d\f,\math{L}_\x\d\f)\geq O(\l)\,.
\end{aligned}\ea}
Using the explicit expression of the line element, we have
\ba\begin{aligned}\label{xy}
R^{ri}_{vi}=y=\frac{2f \pd_v \m-\m \pd_v f}{2r \m^3}\,,\ \ \ R^{ij}_{ij}=x=\frac{\m^2-f}{r^2\m^2}\,,
\end{aligned}\ea
{ with $i\neq j$, where the indices $i,j$ denote the coordinates $\{\q_1,\cdots, \q_{n-2}\}$. Considering the result that $\d(T^\text{EM})_v^r=0$ on the background geometry, the optimal condition of the first-order inequality gives $\d G^r_v=0$ on $\S_h$. } Using Eq. \eq{xy}, we have
\ba\begin{aligned}
G^r_v&=-\sum_{k=1}^{k_\text{max}}\frac{\a_k}{2^{k+1}}\d^{r a_1b_1\cdots a_k b_k}_{v c_1 d_1\cdots c_k d_k}R^{c_1 d_1}_{a_1 b_1}\cdots R^{c_k d_k}_{a_k b_k}\\
&=-\sum_{k=1}^{k_\text{max}}\frac{k \a_k (n-2)!}{(n-2k-1)!}x^{k-1}y\,,
\end{aligned}\ea
which gives
\ba\begin{aligned}
\d G^r_v&=-\sum_{k=1}^{k_\text{max}}\frac{k \a_k (n-2)!}{(n-2k-1)!}\left[x^{k-1}\d y+(k-1)x^{k-2}y\d x\right]_{r=r_h}\\
&=\sum_{k=1}^{k_\text{max}}\frac{k \a_k (n-2)!\pd_v\d f}{2(n-2k-1)!r^{2k-1}_h}\\
&=\frac{2S'(r_h)}{\W_{n-2}}
r_h^n\pd_v\d f\,.
\end{aligned}\ea
For the case with $S'(r_h)>0$, the optimal condition of first-order perturbation inequality can be shown as $\pd_v \d f(r_h,v)=0$. { Considering the assumption that the perturbation vanishes on the bifurcation surface, i.e., $\d f(r_h, v_B)=0$, we can see that the optimal condition also implies $\d f(r_h, v)=0$ for any $v$.}

Then, we turn to evaluate the gravitational part of $\math{E}_h$. Under the optimal condition, we can get
\ba\begin{aligned}
\d R^{ri}_{vi}=\d R^{ij}_{ij}=0\,.
\end{aligned}\ea
Then, it is not difficult to check that
\ba\begin{aligned}
\d_1\left(\m P^{acrd}\right)\grad_a\d_2g_{cd}&\propto \d_1\m(\pd_v \d_2 \m+\pd_r \d_2f)\\
&+\d_2\m(\pd_v \d_1 \m+\pd_r \d_1f)+O(\e)
\end{aligned}\ea
on $\S_0$, { where we have used the optimal condition that $\d f(r_h,v)=0$ and the fact that the background spacetime is a nearly extremal black hole such that $f'(r_h)=O(\e)$. From the above expression, we can see that the variational operators $\d_1$ and $\d_2$ are symmetric, which indicate that
\ba\begin{aligned}
\int_{\S_h}\bm{\w}^\text{LL}(\f, \d_1\f, \d_2\f)=O(\e)\,.
\end{aligned}\ea
From these results, we have
\ba\begin{aligned}
\math{E}_h\geq O(\l, \e)\,.
\end{aligned}\ea
}Then, The second-order perturbation inequality \eq{varid22} reduces to
\ba\begin{aligned}
\d^2M-\F_H\d^2Q-\frac{4\p\d Q^2}{(n-3)\W_{n-2}r^{n-3}_h}\geq O(\l,\e)\,.
\end{aligned}\ea

Next, we evaluate the value of $h(\l)$ under the second-order approximation of perturbation for the case with $S'(r_h)\geq 0$. Under the second-order approximation of $\l$, we have
\ba\begin{aligned}
&h(\l)=f(r_m)+\l \d f(r_m)\\
&+{ \frac{\l^2}{2}}\left[\d^2 f(r_m)+2\d r_m \d f'(r_m)+\d r_m^2 f''(r_m)\right]+O(\l^3)
\end{aligned}\ea
For the zero-order term of $\l$, we have
\ba\begin{aligned}
f(r_m)=-\frac{1}{2}\e^2r_h^2 f''(r_h)+O(\e^3)\,.
\end{aligned}\ea
{ For the first-order term, according to Eq. \eq{fprm} as well as the definition of the blackening factor, we can get
\ba\begin{aligned}
&\d f(r_m)=-\frac{16\p }{\W_{n-2}\math{T}(r_m)}\left(\d M-\frac{4\p Q\d Q}{(n-3)\W_{n-2}r_m^{n-3}}\right)\\
&= -\frac{4\p}{S'(r_h)}\left[\d M{ -}\F_H \d Q-(n-3)\e \F_H \d Q\right]+O(\e^2)\\
&=\frac{4\p(n-3)\e \F_H \d Q}{S'(r_h)}+O(\e^2)
\end{aligned}\ea
in which the last step used the optimal condition of first-order perturbation inequality.} Variation of Eq. \eq{fprm} gives
\ba\begin{aligned}
&\d^2f(r_h)=-\frac{4\p}{S'(r_h)}\left[\d^2M-\F_H\d^2 Q-\frac{4\p\d Q^2}{(n-3)\W_{n-2}r_h^{n-3}}\right]\,,\\
&\d f'(r_h)=-\frac{4\p (n-3)\F_H\d Q}{r_h S'(r_h)}\,,\\
&\d r_m=\frac{4\p (n-3)\F_H\d Q}{r_h f''(r_h)S'(r_h)}+O(\e)\,.
\end{aligned}\nn\\\ea
Combing the above results and together with the second-order perturbation inequality, we can further obtain{
\ba\begin{aligned}
h(\l)&\leq -\frac{[4(n-3)\p\l\F_H \d Q-r_h^2\e S'(r_h) f''(r_h)]^2}{2 r_h^2 S'(r_h)^2 f''(r_h)}\\
&+O(\l^3,\l^2\e,\cdots)\,.
\end{aligned}\ea
Substituting $f''(r_h)=f''(r_m)+O(\e)$ into the above identity, we have
\ba\begin{aligned}
h(\l)&\leq -\frac{[4(n-3)\p\l\F_H \d Q-r_h^2\e S'(r_h) f''(r_h)]^2}{2 r_h^2 S'(r_h)^2 f''(r_m)}\\
&+O(\l^3,\l^2\e,\cdots)\,.
\end{aligned}\ea
Note that $f''(r_m)\geq 0$ for the nearly extremal black holes. Under the second-order approximation of perturbation, where the $O(\l^3,\l^2\e,\l\e^2,\e^3)$ term and higher-order term are neglected, we have $h(\l)\leq 0$.} This result shows that the nearly extremal static charged black hole cannot be overcharged in the above perturbation process under the second-order approximation. Therefore, the WCCC is valid for the Lanczos-Lovelock-Maxwell gravity with the condition $S'(r_h)> 0$.

\section{Conclusion and Discussion}\label{sec6}

In this paper, we tested the WCCC in the nearly extremal static charged black holes in Lanczos-Lovelock-Maxwell gravity by using the new version of the gedanken experiments proposed by Sorce and Wald \cite{SW}. After assuming that the perturbed spacetime finally settles down to a static state and the matter fields satisfy the null energy condition, we found that the black hole horizons cannot be destroyed for the case with $S'(r_h)>0$ under the second-order approximation of perturbation, and however it can be overcharged for the case with $S'(r_h)\leq 0$ even though we only consider the first-order approximation. This indicates that the WCCC will give a constraint on the Lovelock gravitational theories. As an example, we can consider a third-order Lovelock gravity with the nonvanishing parameters $\a_0, \a_1=1, \a_2$ and $\a_3$. If we demand the condition $S'(r_h)>0$ is satisfied for all of the extremal black holes in Lanczos-Lovelock gravity, we can get a constraint of the parameters and it can be expressed as
\ba\begin{aligned}
\a_3>0\,,\quad \a_2<(n-5)(6-n)\sqrt{\frac{3(n-7)!\a_3}{(n-3)!}}\,.
\end{aligned}\ea
That is to say, WCCC will play a natural role to constraint a gravitational theory.
{
Next, we would like to connect the condition $S'(r_h)> 0$ to the thermodynamical properties of the nearly extremal black hole in the Lanczos-Lovelock-Maxwell gravity. The specific heat of the black hole is
\ba\begin{aligned}
C_Q&=T\left.\frac{\pd S(r_h)}{\pd T}\right|_Q\\
&=\frac{f'(r_h)}{2\p} S'(r_h)\left.\frac{\pd r_h(T,Q)}{\pd T}\right|_Q.
\end{aligned}\ea
Then, from the identities $f(r_h, M, Q)=0$ and $T=(1/4\p)\pd_r f(r_h, M, Q)$, we can further obatin
\ba\begin{aligned}
\frac{\pd r_h(T, Q)}{\pd T}=\frac{\pd_M f}{-f'(r_h)\pd_r\pd_Mf+\pd_M f f''(r_h)}\,.
\end{aligned}\ea
Using the result $f'(r_m)=0$ with $r_m=r_h(1-\e)$, we can further obtain
\ba\begin{aligned}
f'(r_h)=\e r_h f''(r_m)= \e r_h f''(r_h)+O(\e^2)\,.
\end{aligned}\ea
Then, we have
\ba\begin{aligned}
C_Q=\frac{\e r_h S'(r_h)}{2\p}+O(\e^2).
\end{aligned}\ea
That is to say, under the first-order approximation, $C_Q$ has the same signature with $S'(r_h)$. When $S'(r_h)$ is negative, the spacetime is thermodynamically unstable. This result indicates some deep connections between the WCCC and thermodynamics of the black holes.

Moreover, from the viewpoint of the loop quantum gravity, the entropy of the black holes represents the number of the microstates of the gravity on the horizon. Therefore, the condition $S'(r_h)>0$ implies that there would be more zoom for microstates when the black hole gets bigger. In this perspective, this condition may be reasonable for our universe.

Finally, it should be mentioned that we only consider the perturbation which has strictly spherical symmetry. The story may be changed if we consider more general cases with small asymmetries. These further investigations would left for our future work.}

\section*{Acknowledgement}
Jie Jiang  is supported by the National Natural Science Foundation of China (Grants No. 11775022 and
11873044). Ming Zhang is supported by the Initial Research Foundation of Jiangxi Normal University with Grant No. 12020023. The authors are grateful to the anonymous referees for their useful comments which have significantly improved the quality of our paper.

\end{document}